\documentclass[prb,twocolumn,showpacs,preprintnumbers,amsmath,amssymb]{revtex4}
\usepackage{graphicx}% Include figure files 
\usepackage{dcolumn}% Align table columns on decimal point 
\newcommand{\beq}{\begin{equation}} 
\newcommand{\eeq}{\end{equation}} 
\newcommand{\beqa}{\begin{eqnarray}} 
\newcommand{\eeqa}{\end{eqnarray}}

\newcommand{\ra}{{\rightarrow}}

\begin{document} 
\title{Optical sum rule anomalies in the cuprates: interplay between strong correlation and electronic bandstructure}
\author{A. Toschi$^{1}$, M. Capone$^{2,3}$}  
\affiliation{$^1$ 
Max Planck Institut f\"{u}r Festk\"{o}rperforschung, Heisenbergstrasse, 1, D-70569, Stuttgart, Germany}.
\affiliation{$^2$ SMC, CNR-INFM and 
Dipartimento di Fisica, Universit\`a di Roma ``La Sapienza'',
Piazzale Aldo Moro 2, I-00185 Roma, Italy}
\affiliation{$^3$ISC-CNR, Via dei Taurini 19, I-00185 Roma, Italy} 

\begin{abstract}
We theoretically analyze some of the anomalies of the optical sumrules in the 
high-temperature superconductors. In particular we address the particularly strong dependence on temperature of the sumrule in the normal state.
Both electron-electron correlations and the presence of a Van-Hove singularity have
been shown to enhance such a dependence. Here we consider both effects simultaneously
by means of Dynamical Mean-Field Theory for a two dimensional Hubbard model with
realistic parameters for different cuprates, and we find that the two effects are not
cooperative, as they appear to compete one another in the region of parameters relevant
for the experiments.

\end{abstract}

%%%%%%%%%%%%%%%%%%
\date{\today} 

\maketitle 

\section{Introduction}

The phenomenon of high-temperature superconductivity in 
strongly correlated materials such as the cuprates, still represents  one of
 the most challenging topic in condensed matter physics twenty 
years after the discovery. A reason for the difficulty in dealing with the 
cuprates 
is certainly the richness and complexity of their phenomenology, which gives
 rise to many 
competitive explanations, making it extremely hard to 
identify the relevant physical processes underlying the outstanding 
properties of these compounds.
In this light, very important information would be inferred extracting 
from the whole body of experimental observations, the data measuring general and fundamental 
properties (e.g., thermodynamic quantities and sum rules),
focusing as much as possible on 
ubiquitous (material independent) aspects.

One important example, in the context of the infrared spectroscopy 
experiments, is represented by the analysis of the 
optical sumrule (SR)\cite{revshac}. A big effort has been devoted in the 
last few years 
to the  evaluation of the frequency integrals of the optical conductivity 
$\sigma(\omega)$ in several cuprates (e.g.,  $Bi_2 Sr_2 Ca Cu_2 O_{8+x}$ 
(BSCCO)\cite{vdm,bont,deut,carbone}, $La_{2-x} Sr_x Cu O_4$ 
(LSCO)\cite{michele}, $Y Ba_2 Cu_3 O_{6+x}$ (YBCO)\cite{homes,boris}, and more recently also $Hg$-based cuprates\cite{vdmHg}), and to the study of their behavior as a function of 
temperature and doping. One of the most general results emerging from these studies is the strong 
temperature dependence of the optical spectral weight (or partial optical sum rule) defined as
\begin{equation}
\label{eq:defW}
W_{opt}(T,x) = \int_{-\Omega_C}^{\Omega_C} d\omega \; \sigma(\omega,T,x),
\end{equation}
where $\Omega_C$ is an upper cut-off, whose role will be discussed in the following.

More in detail, leaving aside the behavior of the superconducting phase which
establishes below the critical temperature $T_c$, where
even the sign of the temperature variation of $W_{opt}$ depends on 
doping\cite{deut,carbone},  
the main results of the infrared estimate of the partial SR in the normal phase ($T > T_c$) of the cuprates are essentially two:
(i) A strong enhancement of the 
$W_{opt}(T,x)$ is observed when $T$ is reduced. Such enhancement is generally 
quadratic in temperature and is not strongly doping dependent; (ii) The 
extrapolated  spectral weight at $T=0$, $W_{opt}(0,x)$ displays, instead, 
a clear doping dependence, increasing monotonically with $x$.
The qualitative results do not depend significantly on the value of the cut-off
$\Omega_C$ and they hold in particular when it reaches the plasma frequency 
(around 1 eV).

In this work we present a detailed analysis of the behavior of the optical 
sumrule based on Dynamical Mean-Field Theory, a non perturbative many-body approach
 which allows for an accurate treatment of strong correlation effects, which 
turn out to be crucial to account for the experimental observations. 
We base our analysis on the previous work of Ref. \onlinecite{prl}, 
which we extend by including realistic two-dimensional bandstructures with 
Van Hove singularities.

The paper is organized as follows:  In Sec. II we discuss previous results on non
interacting models, and on the qualitative effect of correlations. Sec. III
contains a brief introduction to DMFT and our solution. Sec. IV is devoted to the
discussion of our DMFT results, while a simple interpretation of the same results
is presented in Sec. V. Sec. VI is dedicated to concluding remarks.

\section{Non interacting models}

Most of the interest in the optical SR comes from its relation with the 
kinetic energy\cite{hirsh,norman}. 
The most straightforward identification is actually limited to tight-binding 
(TB) models with nearest-neighbor hopping, where, both in the presence and in the absence of electron-electron
interactions we have
\begin{equation}
\label{eq:TB}
W_{opt}^{TB}(T) = - \frac{\pi e^2 a^2}{2V} E_{kin} (T),
\end{equation}
where $a$ is the lattice constant, $V=a^2$ the cell volume, and $E_{kin}$ 
the kinetic energy of the carriers.
The relation holds if the cutoff $\Omega_C$ 
is chosen large enough to contain the 
whole optical spectrum. 
For a non interacting system this would correspond to integrate up to the 
``plasma frequency'' of the model. We will see briefly that this identification
can be significantly modified by the inclusion of next-nearest (and further) 
neighbor hopping.

Eq.~(\ref{eq:TB}), connecting the SR to the kinetic energy, leads to a
 potential trivial explanation of point (i). The $T^2$ behavior can indeed 
be recovered through a simple Sommerfeld 
expansion of (\ref{eq:TB}), which gives indeed $W_{opt}^{TB} =
 W_{opt}^{TB}(0)-BT^2$, with $W_{opt}^{TB}(0) \propto t$ 
and $B \propto t^{-1}$ (for a flat DOS of bandwidth $8t$, e.g., 
$B=e^2 \pi^3/48t$, see Ref. \onlinecite{lara}).

On the other hand, as pointed out in Refs. \onlinecite{lara,prl,michele},
this one-parameter model can not account simultaneously for the experimentally
measured values of
$W_{opt}(T=0)$ and of $B$. In particular, realistic hopping parameters are 
not compatible with the size of the observed temperature variation of 
$W_{opt}$.
The experimental value of $B$ could be recovered only invoking a value 
of $t \sim 20 $ meV smaller {\sl by more than one order of magnitude} than the 
values determined either theoretically by means of bandstructure 
calculations\cite{eva}, or experimentally through photoemission
 measurements\cite{arpes}.
The simple nearest-neighbor non interacting model appears even poorer
 when considering the behavior of the zero temperature sumrule (ii).
First, the above mentioned value $t \sim 20$ meV is totally incompatible 
with the
experimental values of $W_{opt}(0,x)$  (which ranges from 
$200 \div 500$ meV). Second, the simple model predicts  $W_{opt}(0,x)$ 
decreasing with the doping level, 
just opposite to the observations.

A first natural step to heal the inadequacy of the above nearest-neighbor
 model is to include
a more realistic bandstructure.  In particular as we already mentioned, 
Eq. (\ref{eq:TB}) is valid only for pure  nearest-neighbor hopping. When 
releasing this restriction, 
the generalization of Eq. (\ref{eq:TB}) reads
\begin{equation}
\label{eq:TB2}
W_{opt}^{TB}(T,x)= \frac{\pi e^2}{N_k V} \sum_{ {\bf  k}, \sigma} 
\frac{\partial^2  \epsilon_{{\bf k}}}{\partial {\bf k}_x^2 } 
n_{\sigma}(\epsilon_{\bf k})
\end{equation}
where the sum  is performed over all the $N_k$  momenta  ${\bf k}$ of 
the first Brillouin zone, $\epsilon_{{\bf k}}$ is the dispersion and 
$ n_{\sigma}(\epsilon_{\bf k})$ the occupation number for a 
given ${\bf k}$ state.

It has been pointed out in Ref. \onlinecite{vdm07} that in the 
non-interacting case  the inclusion
of the next-nearest neighbor hopping term $t^{'}$ can determine remarkable 
changes in the 
above picture. More specifically, for realistic values of the hopping 
parameters, remarkable
differences between the kinetic energy and the optical spectral weight 
behavior (defined in Eq. (\ref{eq:TB2})) appear, mainly for dopings close to
 the two-dimensional
Van Hove singularity (VHS): while the temperature dependence of $E_{kin}$ 
is only weakly affected by the doping level, $W_{opt}^{TB}$ is more 
sensitive to 
the VHS, whose proximity
determines a stronger temperature dependence.
Although a stronger $T$-dependence of the spectral weight goes certainly 
in the direction of the experimental evidence, many inconsistencies remain: 
First of all 
the enhancement of the $T$-dependence of $W_{opt}^{TB}$ is still 
 not enough to account for the experiments; secondly for some 
(and realistic)  values of the hopping parameters the results can even show 
a change of sign in the temperature variation of $W_{opt}^{TB}$, 
which has never been observed experimentally. Finally -as already noted 
in Ref. \onlinecite{vdm07}- the doping dependence of $W_{opt}^{TB}(0,x)$ remains opposite to the experimental 
data.

The failure of these simple noninteracting models to capture the experimental
 behavior
can not be surprising in light of the unquestioned role of electronic 
correlations in 
the cuprates. In  Ref. \onlinecite{prl} we have shown that the inclusion 
of correlations
determines indeed a huge step ahead in the understanding of experimental data.
A strong suggestion to proceed in this direction comes also from point (ii)
(see Introduction), because the monotonically increasing values of $W_{opt}(0,x)$ would find a very natural explanation in a strongly  correlated scenario, 
where the electronic mobility is minimal at half-filling and it increases with doping. In Ref. \onlinecite{prl} it is demonstrated how the presence of strong 
interactions can actually determine a {\it separation of the energy scales} controlling $W_{opt}(0,x)$ and its temperature dependence. 
More specifically, if one performs the frequency integral of $\sigma(\omega)$ up to a cut-off which includes both the Drude and  the Mid Infrared (MIR) contribution, mimicking the experimental situation, a strong $T$-dependence of $W_{opt}$, and a qualitatively correct behavior of $W_{opt}(0,x)$ are obtained.

On the other hand,  some discrepancy with the experimental observations is
present also in the data of Ref. \onlinecite{prl}: the DMFT data, which are
 computed for a Hubbard model with a semicircular dispersion which has no VHS,
predict a stronger temperature dependence of $W_{opt}(T,x)$ at small doping 
than for the overdoped compounds, as a result of a larger distance from the
 Mott transition, in contrast with experiments, where this effect is not seen, 
and even an opposite behavior occurs in BSCCO\cite{deut, carbone}.
One of the reasons for such a discrepancy is that DMFT neglects spatial correlations. 
The effect of this neglect is expected to be stronger for smaller dopings\cite{civelli}.
A consequence of the lack of spatial correlation is the vanishing quasiparticle renormalization factor $Z$ (renormalization of the coherent electronic  bandwidth) when $x \to 0$ in contrast with photoemission data, in which $Z$ is finite for any doping. 

Thus, a crucial step for a proper analysis of the small doping region is to consider cluster extensions of DMFT able to capture at least short-range 
correlations (such as Cellular DMFT\cite{cdmft} and DCA\cite{dca}), where the Mott transition can take place with a finite $Z$. Present Cluster-DMFT studies have been mainly dedicated to the superconducting phase, which is not addressed in this paper\cite{Haule,Jarrel}.

In this work we do not address the cluster extensions of DMFT, while we try 
to supplement
our previous single-site DMFT analysis by including the effects of a more 
realistic
two-dimensional bandstructure displaying a VHS, which, as we commented above, can introduce
remarkable effects for non interacting systems. Keeping in mind the limitations of single-site
DMFT, we will not consider extremely small dopings, where non local correlations will become
crucial.

We notice in passing that the $T^2$ behavior of the SR is not limited to  noninteracting systems, but it is rather characteristic of all Fermi-liquid systems, 
either weakly or strongly correlated. The actual coefficient is generally a 
function of $Z$ and of the imaginary part of the self-energy (proportional to  $T^2$ in a Fermi-liquid).
What correlation effects and interaction with bosonic modes can change are 
on one hand the actual values of the $T=0$ value and of the coefficient $B$,
as we have already found, and on the other hand they can introduce coherence 
scales which set the limit under which  the system behaves like a Fermi liquid. 
Beyond that scale the temperature behavior can appear different from quadratic, 
but such a deviation is not necessarily the consequence of deep changes in the 
groundstate properties.

\section{DMFT for the two-dimensional Hubbard model}

In this paper we consider the two-dimensional Hubbard model, i.e., 
\begin{eqnarray}
\label{hubbard}
{\cal H} &=& \sum_{ij\sigma} t_{ij} c_{i\sigma}^{\dagger} c_{j\sigma} 
+U\sum_{i}\left ( n_{i\uparrow}-{\frac{1}{2}}\right )
\left ( n_{i\downarrow}-{\frac{1}{2}}\right )+\nonumber\\
& & -\mu\sum_i (n_{i\uparrow}+n_{i\downarrow}),
\end{eqnarray} 
where $c_{i\sigma}(c^{\dagger}_{i\sigma})$ are annihilation(creation) operators
for fermions of spin $\sigma$ on site i, $n_{i\sigma} = c^{\dagger}_{i\sigma}
c_{i\sigma}$, and the sum in the first term includes nearest-neighbors
(NN) and next-to-nearest neighbor (NNN) hopping processes, whose amplitudes are given  by $-t$ and $t'$ respectively .

Our choice of the parameter appearing  in the Hamiltonian (\ref{hubbard}), 
namely $t,t', U$  aims to the closest contact with the 
cuprate properties.
Specifically, we concentrate on two of the most studied cuprates, i.e.,
 BSCCO and LSCO, for which a certain  agreement about the estimates of the 
hopping parameters has been reached both on the theoretical 
(density-functional theory calculations) and the experimental side (photoemission). 
In particular, for both BSCCO and LSCO the NN hopping parameter $t$ is estimated around 
$400 meV$, while the different crystal structure of their unit cells reflects
in different NNN hopping term: LDA calculations predicts a $t'= 0.17 t$ for 
LSCO, whereas  larger values of $t'$ are estimated for BSCCO 
$t'=(0.25 \div 0.30) t$.   
Let us just note that, in the non-interacting case, these values correspond
a VHS located at lower energy with respect to the half-filling chemical potential: 
In the case of $t'=0.17 t$ (LSCO), the VHS would be crossed at a doping level 
slightly below $x=0.15$, while for $t'=0.30 t$ (BSCCO) the VHS would not be reached even for the 
higher doping levels relevant for the cuprates, being located roughly at $x=0.29$. 

The choice of the value of the repulsive term $U$ is certainly less obvious: 
Although the relevance of electron-electron correlations is widely, if not universally,  accepted as a key element in the properties of the cuprates, it is difficult to estimate precisely its value.
In this paper, we have chosen $U= 12 t$, consistently with Ref. \onlinecite{prl}, which allows for a direct comparison with the results for the simpler semicircular DOS used in that paper, and it allows for a clear separation between the MIR and the Hubbard contributions in the optical sum rule (see next section). 
More generally, the choice of $U= 12 t$ is also guided by experimental evidence from the neutron scattering data in the cuprates, which estimate the antiferromagnetic exchange $J=4t^2/U \sim 140$ meV. A recent DMFT analysis suggests however that a slightly smaller value $U=10 t$ is more appropriate to describe the zero temperature optical spectra and their doping dependence.\cite{Comanac}.

As we anticipated, we use DMFT to solve the Hamiltonian (\ref{hubbard}). 
DMFT is a nonperturbative method that maps a lattice model onto an effective
local model, in which the effect of the neighboring sites on a given site is
expressed through a ``dynamical Weiss field''. The local effective model can
be parameterized by an Anderson Impurity Model (AIM), in which an interacting
site is hybridized with a non interacting bath describing the Weiss field. 
For more details we refer to Ref. \onlinecite{dmft}.
The mean-field is enforced
by a self-consistency condition which relates the Green's function
of the effective model to the local component of the lattice Green's function
of the original model. Namely
\begin{equation}
\label{selfcons}
G(\omega) = \frac{1}{N_k} \sum_{\bf k} \frac{1}{\omega - \varepsilon_{\bf k} + \mu - \Sigma(\omega)},
\end{equation}
where $G(\omega)$ and $\Sigma(\omega)$ are the Green's function and the self-energy of the effective model and $\varepsilon_{\bf k}$ is the bare dispersion on the chosen lattice (from now on, we are setting $a,e=1$) .
The self-consistency condition requires to solve iteratively the AIM until 
Eq. (\ref{selfcons}) is obeyed.
In this work we consider the two-dimensional dispersion $\varepsilon_{\bf k}= -2t (\cos{k_x}+
 \cos{k_y}) +4t'(\cos{k_x}\cos{k_y})$, while in Ref. \onlinecite{prl} we considered a simple
semicircular density of states characteristic of an infinite coordination Bethe lattice. 
We emphasize that the lattice structure enters the DMFT only in the self-consistency condition.

We use exact diagonalization (ED) to solve the AIM. In this method the model can be solved at zero temperature by discretizing the bath function into a small number of levels, which here will be $N_b=7$. The main limitation of the approach is that the spectral properties are those of a finite system, hence the fine details can not be resolved with great accuracy.  This choice is particularly useful for the subject of this work, considering that we are interested in  integrated optical spectra, rather than in their details, and that the temperatures of interest are very small with respect to the energy scales of the Hamiltonian (\ref{hubbard})  (the range of temperature $0< T < 300 K$ corresponds to $0< T < 0.065 t$).

In particular, the very low temperature range allows us to exploit 
the Lanczos algorithm at finite $T$, which  has been developed for DMFT in 
Ref. \onlinecite{lanczosT}. In this scheme we avoid to compute the whole spectrum
of the Hamiltonian, limiting to the relevant low-lying states, allowing for a faster
calculation than the full diagonalization of the matrix. 

\section{DMFT results: spectra and optical conductivity}

Ad discussed in Ref. \onlinecite{prl} and in several precedent works, the
interpretation of the optical spectra in DMFT can be greatly helped by
an inspection of the single-particle spectral function. 
In particular the $k-$integrated spectral function, namely
 the interacting density of states, can be easily computed through the
knowledge of the retarded self-energy $\Sigma_{ret}(\omega)$  on the real axis

\begin{equation}
\label{eq:dos}
N(\omega)= -\frac{1}{\pi} \frac{1}{N_k} \sum_{\bf k} 
\mbox{Im} \frac{1}{\omega+\mu-\epsilon_{\bf k} -\Sigma_{ret}(\omega)}
\end{equation}

\begin{figure}[t!]
\begin{center}
\includegraphics[width=80mm,height=70mm,angle=0]{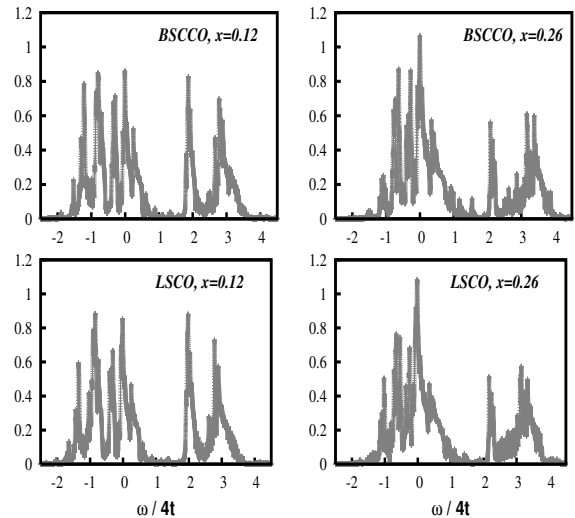}
\end{center}
\caption{\label{fig:dos} Interacting density of states $N(\omega)$  
computed via DMFT at low temperature
($T= 0.02 t$)  for the  case of $BSCCO$ ($t'= 0.3t$) and $LSCO$ 
($t'=0.17t$) for two different values of doping. Note the presence of 
a narrow QP at the Fermi level, very close to the lower Hubbard band, 
while the upper Hubbard band is clearly visible at $\omega= U= 12 t$. }
\end{figure}

In Fig. 
\ref{fig:dos} we plot $N(\omega)$ for $T = 0.02t$, the smallest temperature we considered for two doping levels and for parameters corresponding to the two different materials. Quite generally, the evolution of the density of states is analogous to the Bethe lattice case and contains three features: 
(i) a strongly renormalized Quasiparticle peak (QP) at the Fermi level (of width $\tilde{W} =Z W$, 
where $Z = (1-\partial\Sigma(\omega)/\partial\omega)^{-1}$ is the quasiparticle residue), 
which is basically attached to (ii) the lower Hubbard band (of width roughly equal to the bare bandwidth $8t$), 
and well separated from the upper Hubbard band, whose center is located at $\omega \sim U =12 t$, again with
width close to $8t$.
The weight of the QP clearly increases with the doping level, while the 
spectral gap between the QP and the Hubbard bands appears rather stable in 
the range of doping considered (being of order of $6t \div 8t$). The effect of $t'$ is hardly visible in $N(\omega)$
and seems to affect mainly the shape of the QP, as can be seen by comparing the results for the two values of $t'$ corresponding to BSCCO and LSCO.

Information about the spectral function are particularly useful 
to understand optical spectra due to the simplifications introduced by DMFT
in this regard. The locality of $\Sigma(\omega)$ and of the 
two-particle irreducible vertices in DMFT, together with the odd 
symmetry of the current operator for ${\bf k} \ra - {\bf k}$  
determine in fact the vanishing of all the vertex 
corrections to the current-current paramagnetic kernel\cite{dmft}. As a 
consequence $\sigma(\omega)$ can be computed via the simple ``bubble''
\begin{eqnarray}
\label{eq:sigma}
\sigma(\omega ) &= & \frac{4}{\pi^2 N_k} 
\sum_{\bf k} v_{\bf k}^2 \, \,  \int d\nu \;
Im G_{ret}(\epsilon_{\bf k}, \nu) \nonumber \\ & & 
 Im G_{ret}(\epsilon_{\bf k}, 
\omega+\nu)  \frac{f(\nu)-f(\nu+\omega)}{\omega} 
\end{eqnarray}
where  $v^2_{\bf k} = (\frac{\partial
 \epsilon_{\bf k}}{ \partial k_x})^2$,  
$G_{ret}(\epsilon_{k}, \omega)= (\omega+\mu-\epsilon_{\bf k} 
-\Sigma_{ret}(\omega))^{-1}$ is  the retarded Green function, and finally 
$f(\omega)= (\mbox{e}^{\omega/T}+1)^{-1}$ the Fermi function. For sake of 
simplicity, the normalization of Eq. \ref{eq:sigma} has been chosen so 
that $W_{opt}$ is approaching directly
the value of  $2/N_k \sum_{{\bf k},\sigma}  \frac{\partial^2 \epsilon_{{\bf k}}}{\partial {\bf k}_x^2 } n_{\sigma}(\epsilon_{{\bf k}})$ for  $\Omega_C \rightarrow \infty$.

\begin{figure}[t!]
\begin{center}
\includegraphics[width=80mm,height=70mm,angle=0]{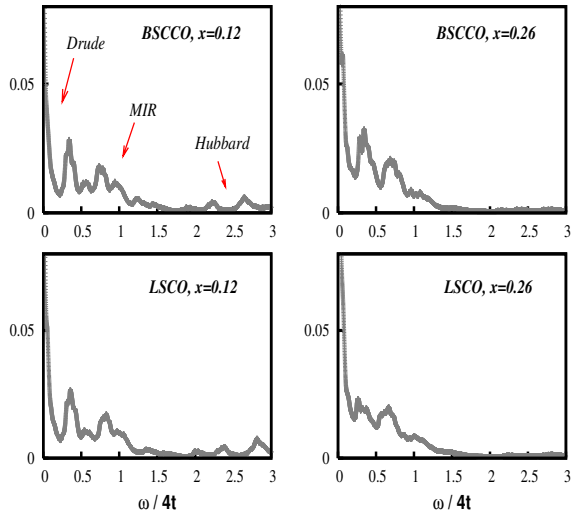}
\caption{\label{fig:sig} 
Optical conductivity $\sigma(\omega)$ computed via DMFT at low temperature ($T= 0.02 t$) for the same parameters of Fig. \ref{fig:dos}. One can immediately notice the narrow Drude peak at low-frequency, a broader MIR contribution at $\omega \sim 4t$ and a small Hubbard term at higher frequencies.}
\end{center}
\end{figure}

Examples of the DMFT results for $\sigma(\omega)$ are shown in Fig. 
\ref{fig:sig} for the same set of parameters of the DOS of Fig. \ref{fig:dos}. One can immediately recognize how the main features of  $N(\omega)$ discussed above reflect in $\sigma(\omega)$. Similarly to the results of Ref. \onlinecite{prl}, the optical conductivity displays (i) a clear Drude peak at low frequencies ($\omega < Z W$), determined by optical transitions occurring at energies within the QP width, then (ii) a Mid Infrared (MIR) bump at $\omega \sim W/2$, related to transitions between the QP and the lower Hubbard band, and finally a high energy contribution at $\omega \sim U$, which is related to transitions involving the upper Hubbard band. 

The rather neat separation between the low energy features (Drude and  MIR)
 and the higher energy Hubbard contribution suggests a value of $\Omega_c \sim 6t \div 8t $ as a natural cut-off to compare (see next section) our calculations with the experimental data.
This choice has the same spirit of choosing the plasma frequency as a cut-off in the experiments in order to select only the contribution from the lowest absorption band. Yet, for our choice of parameters, the cut-off we use is $1.5 \div 2$ times larger than the plasma frequency of BSCCO and LSCO, but using the experimental values in our calculations would result in including only a part of the MIR in the integral, making the analysis less significant, at least in our view.

%\begin{figure}[t!]
%\begin{center}
%\includegraphics[width=80mm,height=70mm, angle=0]{DelWtpri.eps}

%\end{center}
%\caption{\label{} Figure of a typical sigma(omega): Identification of Drude
%MIR and Hubbard contributions  }
%\end{figure}

\section{DMFT results: optical sum rules }

In this section we analyze the results for the optical spectral weight $W_{opt}$ of the Hubbard model through direct numerical integration of the optical conductivity using a cut-off  $8t$ in Eq. (\ref{eq:defW}).
From a theoretical point of view, it is also interesting to compare $W_{opt}$ with the total $SR$ (including in this case also the Hubbard contributions), given by Eq. (\ref{eq:TB2}), where $n_{\sigma}(\epsilon_{{\bf k}}) = -1/ \pi \int d \omega \,  \mbox{Im} G(\epsilon_{\bf k}, \omega) $.

\begin{figure}[t!]
\begin{center}
\includegraphics[width=80mm,height=70mm,angle=0]{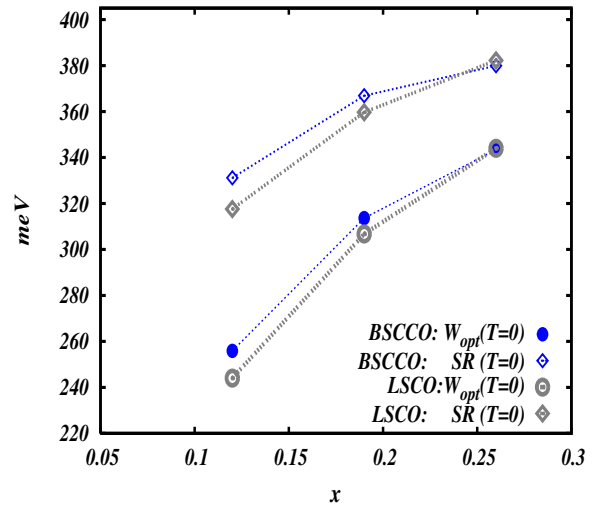}
\end{center}
\caption{\label{fig:W0tpri} Optical spectral weight including Drude and MIR contributions, and total SR computed in DMFT  at $T=0$ as a function of doping for the case of $BSCCO$  and $LSCO$.}
\end{figure}

Even if our focus is the dependence on temperature of these two quantities, a brief analysis of the $T=0$ behavior is necessary before moving to finite temperature.
In Fig. \ref{fig:W0tpri} we report the values of the extrapolated $T=0$ 
values for $W_{opt}$ and SR for three different doping levels for the set of 
hopping parameters which refer to BSCCO and LSCO. 
Both quantities increase as a function of doping, analogously to the Bethe lattice case.
In this respect, as one could have expected, considering  a more
realistic bandstructure dispersion reflects at most in minor corrections
of the DMFT results of Ref \onlinecite{prl}, and in minor differences between the two compounds. This clearly shows that the strong correlation effects play the dominant role in determining the values of both $W_{opt}(T=0)$ and $SR(T=0)$, regardless the shape of the non interacting bands. We notice that, apart from the natural difference (which ranges from $80 \div 40$ meV) due to the contribution of the Hubbard bands, the doping behavior of $W_{opt}(T=0)$ roughly tracks that of $SR(T=0)$, since for the doping considered  the magnitude of the frequency integral in the SR is mostly determined by the MIR and the Drude contributions.

\begin{figure}[t!]
\begin{center}
\includegraphics[width=80mm,height=70mm, angle=0]{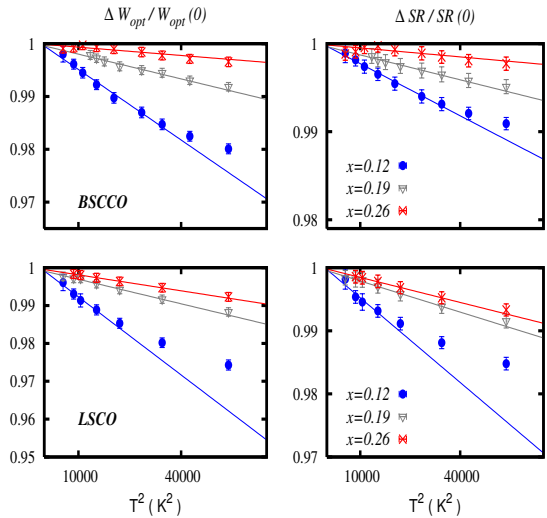}
\end{center}
\caption{\label{fig:DelWtpri} Relative variation with respect to $T=0$ of $W_{opt}$ (left column) and of the total $SR$ (right column) as a function of $T$  for BSCCO (first row) and LSCO (second row). The solid lines are the results of 
the quadratic fit at low temperatures (see text).}
\end{figure}

We turn now to  the temperature dependence of the spectral weight. The results of the DMFT calculations
are reported in Fig. \ref{fig:DelWtpri}, where the ratio between the
value of $W_{opt}(T)$ (and the SR) and its corresponding $T=0$ extrapolation
for the case of BSCCO and LSCO at $x=0.12, 0.19, 0.26$ are shown as a function of temperature.

We see immediately that for both BSCCO and LSCO the relative variation 
$W_{opt}(T)/W_{opt}(0)$ between T=0 and room temperature (in our units $T = 300 K$ corresponds to $0.064 t$)
is larger for the underdoped compounds and constantly decreasing with doping.
For the case of BSCCO such trend is more pronounced than in the Bethe 
lattice case\cite{prl} and it is only partly due to the doping dependence of $W_{opt}(0)$:
A more quantitative analysis through a quadratic fit ($W(T)=W_0-BT^2$, solid lines in Fig. \ref{fig:DelWtpri}) of the data at low $T$ clearly 
demonstrates that the main contribution to this results stems from
a remarkable reduction of the coefficient $B$, which controls the 
low-$T$ behavior. For instance, in the case of BSCCO $B \simeq 21 \mbox{eV}^{-1}$ at $x=0.12$, and it
decreases down to $3 \mbox{eV}^{-1}$ at $x=0.26$. 
In the case of LSCO, the temperature variation at $x=0.19$ is very close to that at $x=0.26$ because
$B$ has a minimum for doping slightly below the VHS.

A second important observation is in order about the temperature 
dependence of the sum rule. With the only exception of overdoped LSCO, in 
the parameter region considered here, the temperature variation of  
$W_{opt}(T)/W_{opt}(T=0)$ is always bigger than that of
 $SR(T)/SR(T=0)$  (for BSCCO by more than a factor $2$):  This results 
is therefore not determined simply by the small difference between $W_{opt}(T=0)$ and $SR(T=0)$: The coefficient of the quadratic fit is always lower in the case of the $SR$ indicating that the 
temperature dependence of the high-energy Hubbard contribution has always the opposite sign with respect to that 
of $W_{opt}$. This should be taken into account when comparing theoretical 
prediction on the spectral weight computed in presence of strong interactions 
with the experimental data. This problem can be faced,e.g., in Cluster DMFT, 
where the  difficulty to evaluate the vertex corrections may suggest to resort to the calculation of the total SR.

The main outcome of these DMFT results is that in presence of a strong correlation the effects of the details of more realistic bandstructure are not very evident in the behavior of the integrated  optical spectral weight. As one clearly sees in Fig. \ref{fig:W0tpri} and \ref{fig:DelWtpri}, the behavior of $W_{opt}(T=0)$ and of its dependence on temperature when the doping level  is enhanced from $x=0.12$ to $x=0.26$ is dominated by the quasiparticle renormalization factor $Z$ which obviously increases with the distance from the Mott transition. In this respect, our DMFT results show that most of the peculiar trends determined by the two-dimensional density of states in a simple non-interacting scheme\cite{vdm07} are washed out by strong interaction effects, so that the main outcomes of Ref. \onlinecite{prl} remain unaltered.

Nonetheless we observe some effects of the more realistic bandstructure also in the presence of strong correlation, and the outcome is not completely obvious.
In particular, one could have expected that, when the chemical potential moves close to a VHS, the largest temperature variation of $W_{opt}$ would have be achieved, because the proximity to the VHS and the strong interaction could have cooperated summed up somehow their effects in making more pronounced the $T$-dependence of $W_{opt}$.  
This is apparently not the case, since the VHS is reached at a doping level slightly below $x=0.15$ for $LSCO$ and larger than $x=0.26$ for BSCCO, and no trace of any enhancement in the $T$-dependence of the spectral weight is found there.
Quite remarkably, instead, one can observe that in BSCCO the weakening of the $T$-dependence  of $W_{opt}$ with increasing doping is more evident than in the Bethe lattice, where no VHS is present, case\cite{prl}, and that in LSCO the change of the temperature variation for doping levels below the VHS (i.e., $x=0.19$ and $x=0.26$) is smaller than expected. 
This clearly  suggest that the proximity to a VHS and the strong interaction effects are partially competitive.
It is worth underlying that no evidence of the change of sign of the temperature variation
when the chemical potential crosses the VHS is found in the LSCO data for $x=0.19$ and $0.26$ as opposed to the 
non interacting case studied in Ref. \onlinecite{vdm07}.

An extremely simple scheme to understand the origin of  these partly 
unexpected  DMFT results will be discussed in the next section.
 
\section{Rescaled temperature dependence: 
A strongly-renormalized Fermi liquid picture}

\begin{figure}[t!]
\begin{center}
\includegraphics[width=80mm,height=55mm, angle=0]{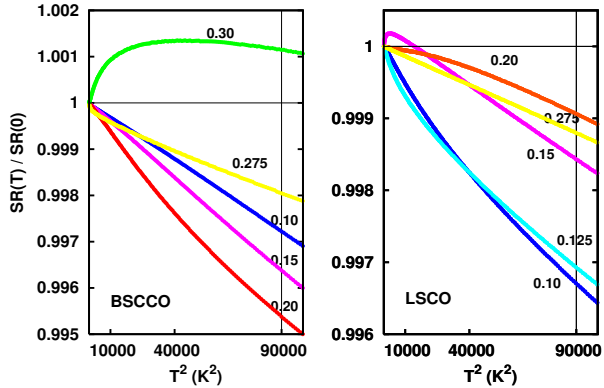}
\end{center}
\caption{\label{fig:Somm} Relative variation with respect to $T=0$ of the SR for a non interacting system with parameters of BSCCO and LSCO.}
\end{figure}

In this section we will present a simple explanation of the surprising interplay between VHS and strong correlations displayed in DMFT. As we discuss below, the main idea behind our explanation is to introduce a simple cartoon for the correlated system, where the SR is given by the value obtained for a non interacting case with bands renormalized by the factor $Z$ obtained in DMFT.
The starting point of our arguments is the temperature dependence 
of the total $SR$ at $U=0$, which was already analyzed in Ref. \onlinecite{vdm07}, and it
is shown in Fig.  \ref{fig:Somm} for the bandstructure parameter considered here and different doping levels.

It is worth noticing that, as it was pointed out in Ref. \onlinecite{vdm07}, the low temperature behavior (which can be obtained also directly  via the Sommerfeld expansion) displays remarkable changes for density close to the VHS. In particular, defining $x_{VHS}$ as the doping corresponding to the chemical potential at the VHS  (i.e., $\mu=-4t'$), when $x$ approaches $x_{VHS}$ from below one observes a strong-enhancement of the slope of the $SR(T)$ at low $T$, while as soon as $x > x_{VHS}$  (e.g., $x=0.30$ for BSCCO, and $x=0.15$ for LSCO) 
the {\it sign} of the slope changes abruptlyto positive, in contrast to natural expectation (and 
 experimental evidence), before changing again sign for  higher doping levels.

One should also note that, as a consequence of the big and rapid changes 
occurring close to the VHS, the temperature behavior of $SR$ can deviate 
rather remarkably from the standard $T^2$ behavior {\it \`a la} Sommerfeld.

As we mentioned in the previous section, however, most of these VHS effects, clearly visible in the non-interacting case, are quite weakened (and, in some cases, they even disappear) in our DMFT results, as a consequence of the strong correlation. To understand the reason for this effect we observe that in the non interacting systems the anomalous behavior of the SR due to the VHS is always limited to a small temperature range, whose size becomes smaller and smaller as the VHS is approached.
If we consider that in presence of the strong interaction the low-temperature physics should be mainly controlled by the strongly renormalized QP at the Fermi level (see again the DOS reported in Fig.~\ref{fig:dos}), one can reasonably expect that correlations determine a reduction (by a factor $Z$) of the size of the low-temperature region which is controlled by the VHS.
 In particular, if the QP is strongly renormalized, as in the case we have considered here (the $Z$ evaluated in DMFT ranges between $0.10$ for BSCCO at $x=0.12$ and $0.25$ for LSCO at $x=0.26$), the effects of the VHS should be limited to temperatures so low to become hardly visible in the experiments and in the temperature range we considered in our DMFT results.

We can try to test this generic idea by mimicking the strongly interacting Fermi liquid physics by simply rescaling by a factor $Z$ the energy scale of the non interacting system. More precisely, the renormalization of the QP peak determines a renormalization of  the coefficient $B$ of the $T^2$ term ($B \to \sim B_0/Z$ because $B$ has the dimensions of the inverse of an energy), this means that the scaled results can be simply obtained by replotting the data of Fig. \ref{fig:Somm} rescaling the $T$ dependence of the SR for a given doping level, with the square root of $Z$ extracted from DMFT (i.e., $T \rightarrow T/\sqrt{Z}$). 

\begin{figure}[t!]
\begin{center}
\includegraphics[width=80mm,height=55mm, angle=0]{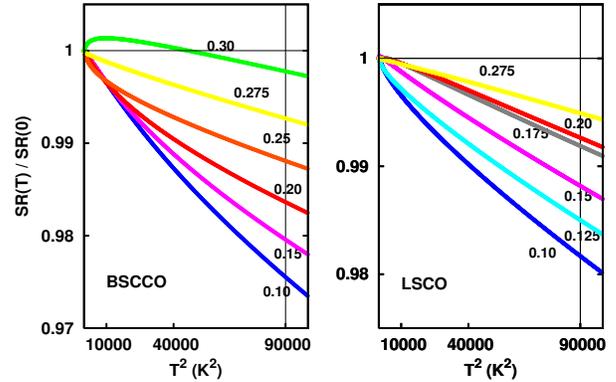}
\end{center}
\caption{\label{fig:Somm_resc} Relative variation with respect to $T=0$ of the SR, after having rescaled the $T$-dependence of the non-interacting model.}
\end{figure}

The results of such rescaling are shown in Fig. \ref{fig:Somm_resc}. Even at first glance, one can notice that the peculiar features associated to the VHS are  much  less evident now, and that the rescaled data resemble in many respects the outcome of the full DMFT calculation. More specifically, the slope of SR(T) does not change sign but in a very low temperature regime, with the exception of the BSCCO at $x=0.30$, which is however above the doping regime of our interest. The same consideration applies to the increasing of the slope, which is observed for $x \to x_{VHS}$  in the non interacting case, and which is now limited to a very tiny low-temperature regime. These results explain why, in the presence of a strong correlation, the effect of the VHS is not only small, but even {\sl opposite} to what one can expect from a Sommerfeld expansion at $U=0$: the deviation from the low-T ``Sommerfeld''  regime occurs really at very low T, partly reducing the effect of the QP renormalization on the overall variation of $SR$ between $T=0$ and $T=300 K$. As a second point, one can note that, apart small deviations, the low temperature behavior of the sum rule can be considered to first approximation quadratic, in agreement with DMFT calculations and with most of the experimental measurements. One should also remark that, after the rescaling, the size of the temperature variation of the SR becomes much closer (some percent going from $T=0$ to $T=300K$) to that of the DMFT data and the experimental observations, while even at the doping of the VHS the overall relative variation of SR does not exceed $ 0.5\%$.

Although one cannot expect this  oversimplified analysis to reproduce exactly all the DMFT results (e.g., it would predict a slightly smaller temperature variation with respect to DMFT for LSCO), it sheds light on the essence of the physics which determines the sum rule behavior in the presence of strong correlation. More precisely it indicates rather clearly that the main features of the integrated optical spectra are determined by the physics of a strongly renormalized Fermi liquid, and the effects of the two-dimensional bandstructure are generally small.

\section{Conclusions}

In this manuscript we presented a DMFT study of the optical sumrules
of the two-dimensional Hubbard model, aiming to understand the interplay of
the strong electron-electron correlations and the properties of the 
two-dimensional density of states, namely the Van Hove singularity.
Even if both correlations\cite{prl} and the VHS singularity\cite{vdm07} have 
been invoked to explain the anomalously large temperature dependence of the 
optical spectral weight, the two effects are found not to be cooperative. 
Our DMFT analysis shows that correlations actually reduce the effect of the 
VHS on the optical sumrules. The mechanism underlying this reduced effect is 
a shrinking of the temperature scale below which the effects of the VHS are 
appreciable. 
When correlations are strong (namely in doped cuprates) this effective 
temperature scale becomes lower than the temperatures accessed in the 
experiments.

This means that the inclusion of a more realistic bandstructure cannot 
alterate the conclusions of Ref. \onlinecite{prl} about the necessity of considering
strong correlations to get the correct order of magnitude of 
observed T-dependence of the optical spectral weight  in the cuprates
(as well as the doping dependence of its $T=0$ extrapolation).  
At the same time, our results imply that the almost negligible doping 
dependence of the temperature variation of $W_{opt}(T)$ which has been 
clearly  observed in several experiments, cannot be explained within a pure 
DMFT calculation, even including the effects of a more realistic 
bandstructure.    

In this respect, non-local corrections beyond the DMFT level  
certainly play a role, since they can introduce different renormalizations for the hopping parameters and the QP weight, and the physics of the Mott transition can become richer, including for example pseudogap features at low temperatures.
The inclusion of such non local effects certainly represent an interesting 
challenge for future studies.

\section*{Acknowledgements}

It is a pleasure to thank G. Sangiovanni, C. Castellani, L. Benfatto, 
A.J. Millis, P. Hansmann, K. Held, L. de' Medici, 
M. Ortolani, P. Calvani for useful discussions. M.C. acknowledges 
financial support of MIUR PRIN 2005, Prot. 200522492.

%Unused bibitems

%\bibitem{chatt} A. Chattopadhyay, A.J.  Millis, and S. Sarma, 
%Phys. Rev. B, {\bf 61}, 10738 (2000).
%\bibitem{arun_qmc} A. Paramekanti, M. Randeria and N. Trivedi,
%Phys. Rev. Lett. {\bf 87}, 217002 (2001). 
%
%\bibitem{science} M. Capone, M. Fabrizio, C. Castellani, and E. Tosatti,
%Science, {\bf 296}, 2364 (2002).
%\bibitem{moeller} G. Moeller, Q. Si, G. Kotliar, M.J. Rozenberg, and 
%D. S. Fisher, Phys. Rev. Lett {\bf 74}, 2082 (1995).
%\bibitem{exe} M. Capone, M. Fabrizio, C. Castellani, and E. Tosatti,
%Phys. Rev. Lett {\bf 93}, 047001 (2004).
\end{document}